\documentclass[11pt, a4paper]{article}

\usepackage{graphicx}
\usepackage{comment}
\usepackage{amssymb}
\usepackage{amstext}
\usepackage{amsmath}
\usepackage{amsthm}
\usepackage{multicol}
\usepackage{pslatex}
\usepackage{xcolor}
\usepackage{multirow}
\usepackage{array}
\usepackage{hyperref}
\usepackage{authblk}
\usepackage{setspace} 

\begin{document}

\title{Graph4J - A computationally efficient Java library for graph algorithms}
\author{Cristian Fr\u asinaru}
\author{Emanuel Florentin Olariu}
\affil{"Alexandru Ioan Cuza University", Ia\c si, Romania}
%\author{Cristian Fr\u asinaru, Emanuel Florentin Olariu}
\maketitle
%\sup{1}\orcidAuthor{0000-0002-5246-7396}
%\email{cristian.frasinaru@uaic.ro}
%\keywords{Graph algorithm, Java library, Runtime performance, Memory usage}

\abstract{
Graph algorithms play an important role in many computer science areas.
In order to solve problems that can be modeled using graphs, it is necessary to use a data structure that can represent those graphs in an efficient manner. On top of this, an infrastructure should be build that will assist in implementing common algorithms or developing specialized ones. 
Here, a new Java library is introduced, called Graph4J, that uses a different approach when compared to existing, well-known Java libraries such as JGraphT, JUNG and Guava Graph. Instead of using object-oriented data structures for graph representation, a lower-level model based on arrays of primitive values is utilized, that drastically reduces the required memory and the running times of the algorithm implementations.
The design of the library, the space complexity of the graph structures and the time complexity of the most common graph operations are presented in detail, along with an experimental study that evaluates its performance, when compared to the other libraries.
Emphasis is given to infrastructure related aspects, that is graph creation, inspection, alteration and traversal. The improvements obtained for other implemented algorithms are also analyzed and it is shown that the proposed library significantly outperforms the existing ones.
%, especially for large inputs.

%All these together form a {\it graph library}.
%The significant runtime improvements obtained for fundamental graph algorithms show that this computationally efficient infrastructure, which takes into account aspects related to graph creation, inspection, alteration and traversal, has a major impact in the overall performance of graph related algorithm implementations.

\section{Introduction}
\label{intro} 

Graphs are mathematical abstractions used to model various theoretical and real-life problems.
Formally, a graph $G$ is a pair $(V, E)$, where $V$ is the set of {\it vertices} and $E$ is a set of ordered or unordered pairs of vertices, representing its {\it edges}.
Due to their simplicity and power of representation, they are used in almost all research areas, 
vertices being the entities of a specific 
domain model and edges describing the relationship between them
\cite{ahuja:1993}, \cite{diestel:2005}, \cite{newman:2010}.
%For an introduction in graph theory the reader can consult \cite{ahuja:1993}, \cite{diestel:2005}, \cite{newman:2010}.
%transportation, social networks, scheduling and planning, 
%network security, computational biology, logistics/planning, psychology, chemistry, and linguistics.

Creating a simple data structure to represent a graph is not exactly a difficult task, regardless of the method of representation and the programming language. 
However, in order to efficiently implement graph related algorithms, that are able to handle large inputs of various types, further efforts are required. 
In addition to creating a graph, methods for inspecting, altering, traversing and other type of common operations are also necessary.
This is the reason why, libraries have been created for most programming platforms, which offer instruments for working with graphs,
usually in the form of a specific application programming interface (API).
In addition to that, many of them also provide  efficient and well-tested implementations of commonly used graph algorithms. 
These libraries are then easy to integrate into various projects,
reducing the programming effort and providing a reliable and high-performance infrastructure.

This paper introduces Graph4J \footnote{\url{https://profs.info.uaic.ro/~acf/graph4j}} \footnote{\url{https://github.com/cfrasinaru/graph4j}}, a library created for the Java programming platform. 
Currently, there are already several production ready libraries for this platform: JGraphT \cite{jgrapht}, JUNG \cite{jung} and Google Guava Graph \cite{guava}. 
While they are great in many aspects, some limitations exist, especially due to performance reasons.
All of them use an object-oriented representation of a graph internals,
where vertices and, in some cases, edges are individual objects, and this comes with penalties regarding performance.
We propose a different, lower-level approach, which does not hinder the modeling ability, 
but instead reduces the memory footprint of the data structures and the running times of the algorithms, sometimes with an order of magnitude.
To prove this, a computational comparison that tests key aspects regarding graph manipulation provided by all these libraries is performed. 
Since JGraphT contains an extensive collection of algorithms, the performance of those implemented so far using the new proposed platform were compared with the corresponding ones in JGraphT.

The remainder of this paper is structured as follows.
Section \ref{relatedWork} discusses related work, enumerating the most well know graph libraries.
The system architecture is presented in section \ref{architecture}, 
where we introduce the graph types that can be defined and how domain specific problems can be modeled using our API.
Section \ref{dataStructures} analyzes in detail the data structure used to represent a graph, its space requirement and the time complexities of the common operations.
The algorithmic layer is discussed in section \ref{algorithms}, both from the user and the developer perspectives.
Section \ref{experiments} covers the computational experiments 
and compares Graph4J with the other libraries - mainly concerning their infrastructure.
We also include here a comparison of some of our algorithm implementations against the corresponding ones in JGraphT.
Finally, section \ref{conclusions} offers the conclusions and the future research and development directions.

%such as transportation, social ne tworks, scheduling and planning, 
%network security, computational biology, logistics/planning, psychology, chemistry, and linguistics.

\section{Related Work}
\label{relatedWork} 

As expected, there are many available software solutions that provide essential tools for any research that involves graph theory. Mathematics-based software, either commercial, 
such as Mathematica \cite{wolfram}, MATLAB \cite{matlab}, Magma \cite{magma}, Maple \cite{maple}, or open-source alternatives such as SageMath \cite{sage}, 
offer a wide range of graph related functions, accessible usually through a dedicated environment, either a Web interface or a native desktop application.

For most programming platforms, there are also graph libraries that provide data structures required for working with graphs and implementations of the most commonly used algorithms.
Written in C++, LEDA \cite{leda} is a comprehensive software platform for combinatorial and geometric computing,
while the Boost Graph Library (BGL) \cite{bgl} comes with a generic programming approach in the context of graphs and also addresses a parallel and distributed model \cite{bgl-par}.
Written in ANSI C, iGraph \cite{igraph} was designed to handle large graphs efficiently, and to be embedded into higher level programming languages (like Python, Perl or GNU R), 
to be used both interactively and non-interactively.
NetworkX \cite{networkx} is a Python package for the creation and manipulation of complex networks.
It includes many standard graph algorithms, in addition to data structures for graphs, digraphs, and multigraphs.
NetworKit \cite{networkit} is also a Python module, with performance-aware algorithms written in C++, which aims to provide tools for the analysis of large networks, with a focus on parallelism and scalability.
The Stanford Network Analysis Platform (SNAP) \cite{snap} is another general-purpose system written in C++ and having Python bindings, that addresses the analysis and manipulation of large networks.
As already mentioned, for the Java programming platform there are three well-established libraries: JGraphT, JUNG and Google Guava.

JGraphT \footnote{\url{https://jgrapht.org}} \cite{jgrapht} contains generic graph data-structures along with an impressive collection of state-of-the-art algorithms.
It is currently used in large scale commercial, non-commercial and academic research projects, 
such as the Apache Cassandra database, the distributed real-time computation system Apache Storm, the Graal JVM, the Constraint Programming Solver Choco, and in Cascading, a software abstraction layer for Apache Hadoop.

JUNG \footnote{\url{https://jrtom.github.io/jung}} \cite{jung} (Java Universal Network/Graph) is a framework for modeling, analysis, and visualization of graphs in Java.
It offers elaborate data structures, including hypergraphs, 
implementations of algorithms from graph theory, data mining, and social networks and an extensive support for graph drawing.

Guava \footnote{\url{https://github.com/google/guava/wiki}} \cite{guava} is the open-sourced version of Google's core Java libraries. 
The package {\tt common.graph} contains a programming interface for modeling graph-structured data and basic support for graph manipulation, traversal and filtering. It does not include other algorithm implementations, 
being in this respect similar to the Java Collections Framework \cite{java}.

\section{The Architecture}
\label{architecture} 

\subsection{Graph Types}
\label{graphTypes} 

A primary goal of the new proposed library is to offer
the ability to define graphs that have any of the following features:
directed or undirected, allowing multiple edges between two vertices or not, allowing edges from a vertex to itself or not,
weighted or unweighted, labeled or unlabeled.

A {\it simple} graph is defined as a pair $G=(V,E)$ where $V=V(G)$ is the set of vertices and $E=E(G) \subset {V \choose 2}$ is the set of edges.
If $e=(v,u)$ (or simply $e=vu$) we say that $v$ and $u$ are adjacent and $e$ is incident to $v$ and $u$. $N_G(v)$ is the set of all vertices adjacent to $v$ (its neighbors) and $d_G(v)=|N_G(v)|$ represents the degree of $v$.
A {\it multigraph} permits multiple edges between two vertices, therefore $E$ is a multiset over $V$ 
and there is a map $m : {V \choose 2} \rightarrow \mathbb{N}$ defining the multiplicity of an edge.
A {\it pseudograph} extends the notion of multigraph by allowing edges between a vertex and itself. Edges are defined over the set $V \cup {V \choose 2}$ and an edge $e$ such that $|e|=1$ is called a loop.
If, instead of being sets, edges are ordered pairs of vertices we obtain {\it digraphs} (directed graphs). In this case $E \subset V \times V$ and its elements are also called arcs (or directed edges).
If $e=vu$ is an arc directed from $v$ to $u$, we say that $v$ is the initial extremity of $e$ (source), $u$ is the final extremity of $e$ (target or sink), $u$ is a successor of $v$, and $v$ is a predecessor of $u$.
The number of successors of a vertex is called its outdegree and the number of predecessors its indegree.
The notions of {\it directed multigraph} and {\it directed pseudograph} are defined in the same way.
By replacing directed edges with simple ones, merging multiple edges and removing self-loops we obtain the {\it support} of any graph variant.

Creating a single data structure to define a generic graph that can be configured depending on what the user wants is not a good design choice 
both because it would be difficult to implement it efficiently and also because it would lead to a confusing programming interface, with all the methods available regardless of the actual graph type. 
On the other hand, creating $2^5$ data types to address every combination of the previously mentioned features would also be inconvenient, 
taking also into account that Java does not allow multiple inheritance at the implementation level.

Taking these aspects into considerations, we have defined a hierarchy of interfaces that describe the main concepts the users would deal with (Figure \ref{fig:GraphHierachy}).

\begin{figure}[htb!]
\begin{center}
\includegraphics[scale=0.6]{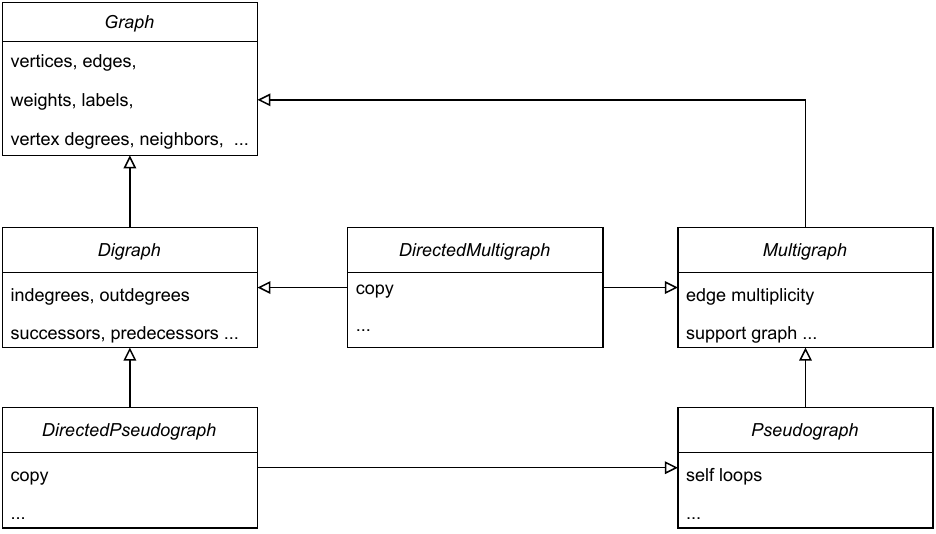}
\caption{The hierarchy of interfaces representing graph data types.}
\label{fig:GraphHierachy}
\end{center}
\end{figure}

The root is represented by the interface {\tt Graph} which defines the methods available to all other types, for handling vertices, edges, iterating, etc. It is also the type used to reference a simple undirected graph. 
%From a formal point of view, a {\tt Graph} is a pair $(V,E)$, where $V$ are the vertices and $E \subset \binom{V}{2}$ are the edges. 
{\tt Digraph} models the concept of simple directed graph, where the edges have a direction from one vertex to another.
%that is $E \subset V \times V$. 
A simple directed graph does not allow multiple edges or self-loops. This data type offers access to specific methods for obtaining the predecessors of a vertex, determining its indegree, etc.
Graphs and digraphs that permit multiple edges, but not self-loops, are represented by the interfaces {\tt Multigraph} and {\tt DirectedMultigraph}, respectively. 
These interfaces have methods for determining the multiplicity of an edge, the set of edges connecting two vertices, etc.
Finally, graphs and digraphs that allow both multiple edges and self-loops are represented by the interfaces {\tt Pseudograph} and {\tt DirectedPseudograph}, respectively.
These interfaces have also specific methods, such as the one for determining the number of self-loops of a vertex.
All of them allow the possibility to make type-safe copies of themselves.

A {\it weighted} graph has primitive numbers associated to its edges, vertices or both, their meaning depending on the context of the problem: distances, probabilities, costs, etc.
In order to closely link a graph to the real-life model it represents, 
both vertices and edges may be put in correspondence to some custom data, represented by user objects 
(e.g. vertices are cities and edges are roads between them or vertices are people and edges represent their friendship relation like in a social network).
We consider such graphs as {\it labeled}.
All the types enumerated so far might be weighted or unweighted, labeled or unlabeled, both for vertices and edges. The methods for handling weights and labels are all declared in the interface {\tt Graph}, 
so are available to all other subtypes.
In order not to waste memory, the data structures that hold information about weights and labels are lazily instantiated, only when a corresponding method is invoked.

All the interfaces that describe graph types have dedicated classes that implement them. These classes are hidden from the user, being declared accessible only at package level. 
Creating actual graph objects is done in a flexible manner, using a {\tt GraphBuilder} that allows the user to configure it step by step.

\subsection{Modeling}
\label{modeling} 
One of the mandatory requirement of our design was to allow both the creation of graphs as mathematical structures - using primitive natural numbers as vertex identifiers, 
and also the creation of graphs where vertices and edges represent domain specific objects - in order to simplify the process of modeling a problem. 
JGraphT, JUNG and Guava libraries address mainly the later variant where objects are used for vertices and, with a single exception provided by Guava, also for edges. 
From the modeling perspective this creates an user-friendly environment, close to the specific problem being solved. 
However, at the algorithmic level, this object-oriented approach is more difficult to handle than a simpler mathematical representation.
As noted in the article describing JGraphT \cite{jgrapht}, in the section dedicated to alternative backends, this flexibility increases memory overhead. 
For this reason, they also offer an implementation with a smaller memory footprint, that uses a third-party library providing collections of primitive values. 
The described performance gains are not significant, since the main data type {\tt org.jgrapht.Graph} is designed to work with reference types.
Considering these aspects, we tried to achieve a better compromise.

Defining a graph using the mathematical model is straightforward.
The following example creates a complete graph with $3$ vertices:
\begin{verbatim}
  Graph g = GraphBuilder.numVertices(3).named("K3")
                .addEdge(0,1).addEdge(1,2).addEdge(0,2)
                .buildGraph();
\end{verbatim}
By default, the vertices of a graph are primitive integers, starting with $0$. However, any non-negative integers can be used as vertices. 
Once a graph is created, it is by default mutable, other vertices and edges can be added or removed as necessary.

In order to address the object-oriented approach, we consider the following flow (Figure \ref{fig:Flow}) that describes the steps required to solve a graph related problem.

\begin{figure}[htb!]
\begin{center}
\includegraphics[scale=0.6]{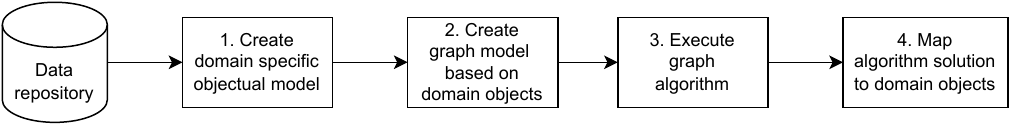}
\caption{The flow of solving an object-oriented graph related problem.}
\label{fig:Flow}
\end{center}
\end{figure}

The input data, once it is read from a repository (database, file, etc.), will be stored in an indexed collection of objects. 
Those indices will represent the vertices of the graph, while the user objects may be attached to them as labels (actually, this is not necessary if the collection of objects is globally accessible).
Creating the model can be done exclusively using user objects, while the resulting graph has still a numerical representation. 
In the following example, each {\tt City} object receives an integer value when it is added into the graph, equal to its index in the array {\tt city}, which will represent the actual vertex. 
\begin{verbatim}
 City[] city = ...; 
 Road[][] road = ...; //read from a repository
 Graph<City, Road> g = GraphBuilder.labeledVertices(city)
   .addLabeledEdge(city[0], city[1], road[0][1])
   .buildGraph();
\end{verbatim}
Getting the object corresponding to a vertex number can be easily done with the method {\tt getVertexLabel(int vertex)}, while {\tt findVertex(E label)} does the opposite, where {\tt E} is the generic type of the labels.
The algorithms will perform on the mathematical model and will benefit from its simplicity.
Once the numerical solution is obtained, it is straightforward to map it back to the user objects, using the methods mentioned before.

\section{Data Structures}
\label{dataStructures} 

\subsection{Graph Structure}
\label{graphStructure}
A graph can be represented in multiple ways. 
An {\it adjacency matrix} $a_{n\times n}$ of a simple graph is defined such that $a_{ij}=1$ if and only if  the vertices $i$ and $j$ are neighbors, where $n$ is the number of vertices. 
Adding or removing edges, and checking vertices adjacency can be performed in constant time, but determining the neighbor list of a vertex has a time complexity of $O(n)$. 
A major drawback of adjacency matrices is their $O(n^2)$ memory requirement, which makes them completely inappropriate for large graphs.
A more space-efficient structure is an {\it adjacency list}. 
For each vertex $v$ is maintained the list of its neighbors. Adding an edge takes $O(1)$ time but removing an edge or checking adjacency are performed in $O(d_G(v))$ time.
However, they are memory efficient, having a space complexity of $O(m)$, where $m$ is the number of edges, and allow the traversal of the entire structure in $O(m + n)$ time, 
making them suitable for large graph implementations.
Other representations exists, such as the {\it incidence matrix} which captures the vertex-edge relationship, 
the {\it cost-adjacency matrix} which is similar to the adjacency matrix but also specifies the edge weights or simply the {\it list of all edges}.

We have chosen to implement the graphs using adjacency lists, as this is both space and time efficient and most of the algorithms will benefit from a simple way to iterate over the neighbors of a specific vertex. 
Algorithms that require other representations, such as an adjacency matrix, will receive them upon request, as we provide methods for creating them when they are needed.
An adjacency list can be implemented in various ways, depending on the specific structures offered by a programming platform. 
We used a lower-level approach based on primitive arrays, unlike the other libraries which mostly rely on {\tt List}s, {\tt Set}s and {\tt Map}s from Java Collections Framework.
From the perspective of writing a simple, elegant code, using collections of objects in order to represent structured data is the preferable way.
For example, a collection of the type {\tt Map<Vertex, Set<Edge>>} might be enough to describe the adjacency lists of a graph, 
with {\tt Vertex} and {\tt Edge} objects encapsulating the additional information (weights, labels, etc).
However, this comes with penalties regarding performance, as we will show in the following paragraphs.

Each Java object allocated on the heap has a {\textit{header}} which contains information used by the Java Virtual Machine (JVM) for locking, garbage collection or the identity of that object \cite{lindholm:2014}.
The size of the header depends on the operating system, but usually in modern JVMs is 16 bytes (on 64 bit architectures).
For example, a primitive {\tt int} takes $4$ bytes of memory while an object of {\tt Integer} type, wrapping the same primitive number, 
takes $16 \text{(object header)} + 4 \text{(content)} = 20$ bytes. This means that using objects to represent graph vertices requires at least $5$ times more memory.
Also, for performance reasons, JVM will {\textit{align}} data.
It means that if we have an object that wraps just one byte, it will not use $8 \text{(object header)} + 1 \text{(content)} = 9 $ bytes of memory on the heap, 
but it will use $16$ bytes as it needs to be aligned to the next $8$ byte boundary.

Representing information using objects is similar to the {\it row-based} form used in most relational databases management systems, 
where sets of rows of the same type form {\textit{tables}}, each table having a number of {\it columns}.
Rows correspond to objects and columns correspond to object properties.
A {\textit{column store}} model~\cite{abadi:2013} "reverses" the orientation of the tables.
It stores data by columns and uses row identifiers in order to access a specific cell of the table.
An important benefit is that column stores (implemented in our case with arrays) are very efficient at data compression, 
since representing information of the same type inside of a column drastically reduces the number of objects and also the data alignment process. 
A more detailed analysis of the benefits of column stores in object-oriented programming can be found in \cite{elprep:2020}.

We have used an array based column store approach in order to represent the main data structure of a graph. This is illustrated in Figure \ref{fig:GraphStructure}.

\begin{figure}[htb!]
\begin{center}
\includegraphics[scale=0.7]{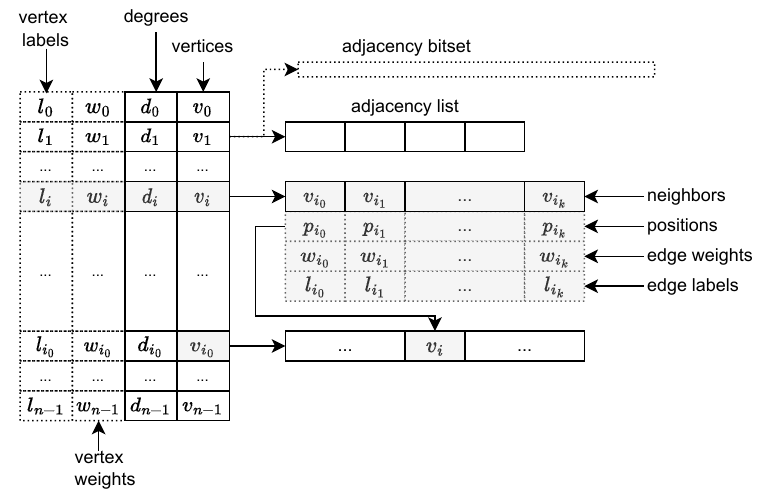}
\caption{Graph data structure. Vertex data is stored in columns (arrays or primitive values) and for each vertex, there are columns storing the information of its incident edges. 
The dotted lines mark the structures that are only created in specific contexts.}
\label{fig:GraphStructure}
\end{center}
\end{figure}

All the arrays in the column store are adjusted dynamically as elements are added to them. 
As a result, the length of the {\tt vertices} array, for example,  might become greater than the actual number of vertices in the graph. 
This may cause a higher memory consumption than the estimated one, if the number of vertices in the graph is not known at creation time. 
We prefer this simple solution over one based on linked lists as it offers positional access to the elements and a fast way to iterate over them.
The adjacency list of a vertex $v=v_i$ contains its neighbors, while the {\it positions} list contains the position of $v$ in the adjacency lists of its neighbors. 
This positions are necessary for undirected graphs since operations regarding an edge $vu$ performed in $v$'s adjacency list, such as setting its weight, label or removing it, must also be performed in $u$'s adjacency list. 
The {\it positions} list allow us to implement the corresponding operations in $O(1)$ time.
The {\it adjacency bitset} is used to test in $O(1)$ if two vertices form an edge. 
For each vertex $v$, an array of $n/64$ {\tt long} values is used to store $n$ bits, one for each vertex of the graph, with those corresponding to the neighbors of $v$ set to $1$ ({\tt long} type is represented on $8$ bytes). 
Used indiscriminately, this would be expensive in terms of occupied memory, with a space complexity of $O(n^2)$. 
This is why we activate it lazily only when the number of elements in an adjacency list exceeds a certain threshold.  
If the degree of a vertex is small, we don't need the bitset since it is easy to iterate over its neighbors. 
If the degree is large, either it is an isolated case or the graph is dense, hence we expect a low number of vertices.
Using a threshold of $\sqrt n$ we have obtained the best results, with a space requirement in accordance with the number of edges 
(if $d_G(v) \ge \sqrt n \; \forall v \in V$ then $2m = \sum_{v \in V} d_G(v) \ge n^{3/2}$). If this structure would take too much memory in certain circumstances, the threshold can be modified by the user.
In addition to the members in Figure \ref{fig:GraphStructure}, we have used additional helpers. 
A {\it vertex-to-index} mapper allows us to identify in $O(1)$ time the index in the vertices array of a particular vertex number. 
This is created automatically in a lazy fashion, only when the vertices of the graph are not the default ones $\{0,\dots,n-1\}$.
If the graph is labeled, and the user requires to identify a vertex number or an edge based on its unique label, we use standard Java {\tt HashMap}s. 
Although very fast, these structures (especially the one for edges) are quite demanding in terms of memory so we activate them only when there is an invocation to a {\tt findVertex(label)} or {\tt findEdge(label)} method.

\subsection{Space Requirement}
\label{spaceRequirement}

The space requirement of all the {\tt Graph} class members is presented in Table \ref{table:graphMembers}. 
In addition to those, there are variables that do not depend on the size of the graph, such as the number of vertices or the number of edges. 
We did not include them when computing the space requirement of a graph, since they represent a negligible amount.

\begin{table}
\small
\begin{center}
\begin{tabular}{ l l l }
{\textbf{Member}}		& {\textbf{Type}} 	& {\textbf{Bytes}} \\
Vertices 				& {\tt int[]} 			& $4n$ \\
Degrees					& {\tt int[]} 			& $4n$ \\
Vertex weights$^*$		& {\tt double[]}		& $8n$ \\
Vertex labels$^*$		& {\tt V[]}				& $4n$ \\
Neighbors 				& {\tt int[][]}			& $4n+8m$ \\
Positions	    		& {\tt int[][]} 		& $4n+8m$ \\
Edge weights$^*$		& {\tt double[][]}		& $4n+16m$ \\
Edge labels$^*$			& {\tt E[][]}			& $4n+8m$ \\
Adjacency bitset$^*$	& {\tt Bitset}			& $m$ \\
Vertex-to-index map$^*$ & {\tt int[]}			& $4n$ \\
Label-to-vertex map$^*$ & {\tt Map<V, Integer>}	& $20n$ \\
Label-to-edge map$^*$ 	& {\tt Map<E, Edge>}	& $\approx 4n+104m$ \\
\end{tabular}
\end{center}
\caption{\label{table:graphMembers}The memory requirement of a {\tt Graph} object. {\tt V} and {\tt E} are generic types for vertex and edge labels - the actual data type is {\tt Object}. 
The members marked with a star are created only if necessary.}
\end{table}

The number of bytes was computed taking into consideration that a primitive {\tt int} requires $4$ bytes, a {\tt double} $8$ bytes, a variable referencing an object $4$ bytes and the header of an object $16$ bytes.
Therefore, a simple undirected graph, with default vertex numbers, has a memory requirement of $16n + 16m$ bytes, 
while an edge weighted graph would take $20n + 32m$ bytes.
These estimations were confirmed during our experiments.
The label-to-edge map is calculated taking into consideration the space required by an {\tt Edge} object. 
An edge has the following properties: {\tt boolean direction}, {\tt int source}, {\tt int target}, {\tt Double weight} and {\tt E label}, that add up to at least $1+4+4+4+4=17$ bytes. 
Setting the weight adds $16+8$ bytes and taking into consideration the header of the {\tt Edge} object, we obtain $17+24+16=57$ bytes. 
A {\tt Map} implementation, such as {\tt HashMap}, will create additional internal objects holding data, 
so it will further increase the required space - the approximation in the table is based on our experiments and it is the only one which was not calculated theoretically.

In the case of digraphs, the adjacency list is split in two: {\it successors} and {\it predecessors}, the occupied memory being virtually the same. 
The degrees array becomes {\it outdegrees} and a new array is required to hold the {\it indegrees}.
The positions in the undirected case are not longer necessary and they are replaced by a similar array holding the positions of a vertex $v$ in the 
adjacency lists of its predecessors.
Although only the list of successors (or predecessors) might be sufficient to represent the digraph, 
we have decided to keep them both as there are important algorithms which could benefit from them, such those for determining maximum flows in transportation networks.

Table \ref{table:digraphMembers} presents the memory requirement of a {\tt Digraph} object (only the differences compared to a {\tt Graph}).
Therefore, a simple directed graph would require $24n + 12m$ bytes,
while a weighted one would take $28n + 20m$ bytes.

\begin{table}
\small
\begin{center}
\begin{tabular}{ l l l }
{\textbf{Member}}		& {\textbf{Type}} 	& {\textbf{Bytes}} \\
Indegrees				& {\tt int[]} 			& $4n$ \\
Outdegrees				& {\tt int[]} 			& $4n$ \\
Successor list			& {\tt int[][]}			& $4n+4m$ \\
Predecessor list 		& {\tt int[][]} 		& $4n+4m$ \\
Predecessor positions	& {\tt int[][]} 		& $4n+4m$ \\
\end{tabular}
\caption{\label{table:digraphMembers}The memory requirement of a {\tt Digraph} object (only the differences compared to {\tt Graph}). Instead of {\tt degrees}, there are {\tt indegrees} and {\tt outdegrees}, 
while {\tt neighbors} and {\tt positions} are replaced by structures specific to directed graphs.}
\end{center}
\end{table}

In the case of {\tt Multigraph}s, there are no other additional members, so the occupied memory will be computed in the same manner.
A {\tt Pseudograph} has a {\tt Map<Integer,Integer>} responsible with counting how many self-loops a vertex has. Its size depends on the number of self-loops, with a maximum estimated at $8n$ bytes.
    
Another observation that has to be made is the one regarding the limits of the graphs that can be represented. 
For all four libraries, the maximum number of vertices is the largest possible integer that can be represented on $32$ bits, 
its exact value being $2,147,483,647$ - that is because the length of an array is of type {\tt int} ($4$ bytes). 
When it comes to edges, Graph4J does not store them in a single collection (it does not store them at all) and it creates them as necessary, using the adjacency lists. 
We declared the number of edges as a {\tt long}, which uses $64$ bits and can hold very large numbers. The other three libraries limit the numbers of edges to the {\tt int} type. 
Actually, in order to find the number of edges in JGraphT or Guava, one has to use constructions like {\tt jgraph.edgeSet().size()} or {\tt guavaGraph.edges().size()}. 
In JUNG, there is the method {\tt getEdgeCount()}, which returns an {\tt int}.
Therefore, Graph4J has the potential to create larger graphs, with billions of edges, provided it has enough memory (a graph with $100$ million vertices and $1$ billion edges takes about $20$ GB).

\subsection{Time Complexity}
\label{timeComplexity}

The time complexity of the default implementation of a {\tt Graph} methods (Table \ref{table:timeComplexity}) are dictated by the structure presented in \ref{graphStructure}, 
which is based mainly on adjacency lists and occasionally on hashtables.

\begin{table}
\small
\begin{center}
\begin{tabular}{ l l l }
{\textbf{Method}}		& {\textbf{Time complexity}}  \\
Adding a vertex				& $O(1)$ \\
Removing a vertex			& $O(1)$ \\
Adding an edge				& $O(1)$ \\
Removing an edge $vu$		& $O(d_G(v))$ / $O(1)$ \\
Testing vertex adjacency	& $O(\sqrt n) / O(1)$ \\
Accessing vertex weights and labels		& $O(1)$ \\
Accessing edge weights and labels 		& $O(d_G(v))$ / $O(1)$ \\
Finding a vertex or edge by its label	& $O(1)$ \\
Finding all vertices by a label	& $O(n)$ \\
Finding all edges by a label	& $O(m)$ \\
\end{tabular}
\caption{\label{table:timeComplexity}The time complexity of operations on a graph ($n = |V(G)|, m = |E(G)|$).}
\end{center}
\end{table}

When adding a new vertex or edge there is an additional overhead required by the array representation. 
If the array is full, a {\it growing} operation must be performed which creates a larger one and copies the corresponding values to it. 
We have used the same approach as in the implementation of {\tt ArrayList} from Java Collections Framework, increasing an array with half of its current size when necessary.
Removing a random edge $vu$ takes $O(d_G(v))$ time since it requires searching the position of $u$ in the adjacency list of $v$. 
The same is true for accessing edge weights and labels. However, most of the time these operations are performed while iterating over the adjacency list and, in this case, their complexity is $O(1)$. 
Note that removing an edge $vu$ while iterating over the neighbors of $v$ is possible in our implementation, 
in the same manner in which a Java {\tt listIterator} permits removing elements out of a {\tt List} while iterating over them.
Testing if a graph contains an edge $vu$ is performed in constant time using a {\tt Bitset} created for $v$, with one bit for every other vertex in the graph. 
As we have mentioned, in order to be space efficient, these bitsets are not created if the degree of $v$ is small (less than $\sqrt n$), in which case we iterate.
Finding vertices and edges by their labels is performed using a {\tt HashMap} in constant time, provided that the {\tt hashCode} and the {\tt equals} methods of the label class are correct.
If the labels are not unique, there are methods that iterate over the whole graph in order to find all vertices or edges that are associated with a particular label.

\subsection{Vertex and Edge Collections}    
Following the same principles of efficiency, we created specialized collections containing vertices or edges (Figure \ref{fig:collections}). 
Unlike the classes in Java Collections Framework, their elements are not objects but primitive values, corresponding to either vertex numbers or edge endpoints. 
Taking into consideration their special nature, we were able to optimize them both in terms of used memory and running time.

\begin{figure}[htb!]
\begin{center}
\includegraphics[scale=0.6]{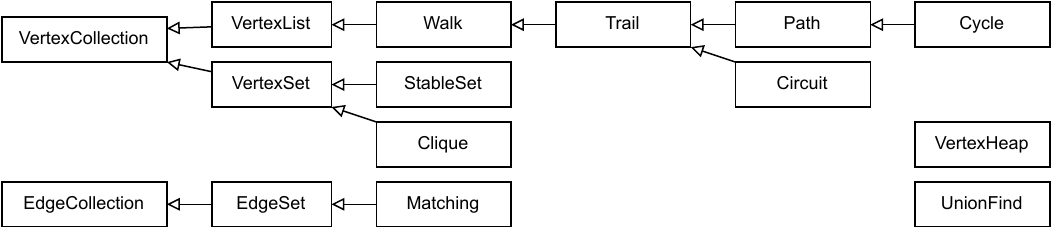}
\caption{Support data structures representing collections of vertices or edges.}
\label{fig:collections}
\end{center}
\end{figure}

A {\tt VertexList} is an ordered collection of vertices, offering precise control over where in the list each element is inserted.
Its implementation is similar to {\tt java.util.ArrayList}, but storing primitive integers instead of object references.
%Unlike sets, lists typically allow duplicate elements.
A {\tt VertexSet} is a collection that contains no duplicate elements and also  offers constant time performance for testing if a vertex belongs to the set or not, the same being true also for {\tt EdgeSet}.
Based on these classes, we have implemented specialized types that describe theoretical graph concepts.
When developing algorithms, these support data structures help in reducing the programming effort and also make the interface most descriptive.
Algorithms will return objects of type {\tt Path}, {\tt Cycle}, {\tt StableSet} or {\tt Matching}, which improves their readability.
In addition, each class that describes a graph concept has a method that checks the validity of its objects, so an algorithm that creates a {\tt StableSet}, 
for example, could simply include an {\tt assert stableSet.isValid()} before it returns the response. In the development phase, when assertions are enabled, this helps early detection of programming mistakes.

\section{Algorithmic Layer}
\label{algorithms}
We designed the algorithmic layer of the library such that it would be intuitive both for the authors of the implementations and for those who use them. 
First of all, the application programming interface must be correlated with the type of graph it handles. All methods in the {\tt Graph} interface should make sense for all types of graphs, and not the other way around. 
For example, the method {\tt neighbors()}, specific to undirected graphs, could be be associated with the successors of a vertex in a digraph. 
On the other hand, methods like {\tt predecessors} and {\tt indegree} make sense for a directed graph, but not for an undirected one. 

In this respect, the author of an algorithm implementation should declare what type of graphs it can handle.
The interface describing the most general type of algorithm is {\tt GraphAlgorithm}. A class implementing this interface would accept any type of graph as input and is expected to perform correctly regardless of its nature. 
For instance, a breadth first search traversal or a shortest path algorithm will simply use a neighbor iterator, defined for all {\tt Graph} objects.
A {\tt SimpleGraphAlgorithm} accepts only simple, undirected graphs, being fit for the acyclic orientation, inspecting connectivity, 
determining if a graph is bipartite, etc. If the input graph is directed, contains multiple edges between two vertices or self-loops, the input of the algorithm will be its support graph.
The interface {\tt DirectedGraphAlgorithm} describes algorithms that are intended only for digraphs, such as topological sort or inspecting strong connectivity. They will accept only inputs of type {\tt Digraph}.

From the user perspective, finding the appropriate algorithm for a specific problem might be challenging, especially if there are more available implementations. 
The algorithm selection process should take into consideration the nature of the problem being solved, requiring specific knowledge about it.
For algorithms that solve the same type of problem, we define a generic interface that describes them, such as {\tt SingleSourceShortestPath}. 
Implementations of that interface could be {\tt DijkstraShortesPath} with or without a heap, {\tt BellmanFordShortestPath}, etc.
At the interface level, there is a default method that selects the most appropriate implementation, based on the input graph.
In this particular instance, a method invocation of type {\tt SingleSourceShortestPath.getInstance(graph)} will produce the implementation which fits best that graph.
If it has negative cost edges, it will select Bellman-Ford, otherwise it might select Dijkstra's and analyze the nature of the graph and determine if it is better to use a heap or not. 
%This feature of our library is still experimental, especially since we do not have too many algorithm implementations yet.

The basic support we provide at the moment addresses the following aspects:
computing the adjacency/cost/incidence matrices, complement and transpose of a graph, line graph, operations such contraction, split, join, union and disjoint union, depth and breadth first search, 
generators for complete graphs, cycles, paths, trees, wheels, stars, tournaments, regular and random graphs and digraphs with various constraints.

The initial release of our library contains also algorithms for: determining graph measures (min/max/average degrees), metrics (distances, eccentricities,  \\ 
girth, radius, diameter, periphery, pseudo-periphery, center), cycle detection, testing connectivity, bi-connectivity and strong-connectivity, identifying bipartite and eulerian graphs, 
topological sorting, acyclic orientation, enumerating maximal cliques, finding single source and all pairs shortest paths, creating minimum spanning trees, greedy coloring, computing maximum flows.
The simplicity of our programming interface and the usage of primitive data structures make the process of writing algorithm implementations straightforward, since their theoretical descriptions found in the literature can be easily followed. 
We verified their correctness using unit testing and comparisons with JGraphT  and the implementations available in the public repository \footnote{\url{https://github.com/kevin-wayne/algs4}} 
that accompanies the book "Algorithms, 4th Edition" \cite{algs4:2011}.

\section{Computational Experiments}
\label{experiments}

\subsection{Methodology}
We mainly tested the infrastructure elements offered by the four libraries, that is: creation, alteration, inspection and traversal of graphs. 
We have used both directed and undirected, dense and sparse, weighted and unweighted graphs. We focused primarily on instances having vertices represented with integer numbers, 
but we also tested the case where user objects are associated both with vertices and edges. We did not test multigraphs and pseudographs as their structure and behavior is very similar to simple graphs.
As Guava and JUNG include none or very little support for classical graph algorithms, we made comparisons only between Graph4J and JGraphT
regarding the performance of some implementations found in both libraries.

Graph4J offers a simple model based primarily on the {\tt Graph} and {\tt Digraph} interfaces, objects of these types being created in a single way, using the class {\tt GraphBuilder}. 
All instances may have weights and labels associated with both vertices and edges. This will not cause performance penalties because their internal structure adapts depending on how they are used. 

JGraphT contains a generic {\tt Graph} interface having a very large number of classes that implement it such as {\tt SimpleGraph}, {\tt SimpleWeightedGraph}, \\ 
{\tt SimpleDirectedGraph}, {\tt SimpleDirectedWeightedGraph}, etc. 
There are also alternative backends, in an optional package, that use third-party libraries, such as {\tt FastutilMapIntVertexGraph}. 
Graph objects can be created either instantiating these classes or using a {\tt GraphTypeBuilder}. Depending on the test, we have selected the most appropriate class, including alternative ones.

Guava offers three unrelated basic interfaces to represent a graph: {\tt Graph}, {\tt ValueGraph} and {\tt Network}, all of them in two versions: mutable or immutable. 
The classes that implement them are hidden by design, the objects being created using {\tt GraphBuilder}, {\tt ValueGraphBuilder} or {\tt NetworkBuilder} classes, according to the case.
{\tt Graph} has edges which are anonymous connections between vertices, with no identity or properties of their own. 
{\tt ValueGraph} has edges which have values, either weights or labels, that may or may not be unique. {\tt Network} has edges which are unique objects, just as vertices. 
%Despite the fact that {\tt Network} is the type that has the same feature as our {\tt Graph}, 
In our tests, we have selected the most appropriate Guava type, depending on the context.

JUNG has also interfaces describing graph types: {\tt Graph}, {\tt UndirectedGraph}, {\tt DirectedGraph}, etc. with a notable mention for {\tt Hypergraph} which is not addressed by the other libraries.
There is no special support for weights, only for user objects that can be associated with vertices or edges. 
Classes that implement these interfaces must be instantiated explicitly, and we have used, depending on the context, either {\tt UndirectedSparseGraph} or {\tt DirectedSparseGraph}. 
The names of the classes suggest they are focused on sparse graphs, but there are no other alternative implementations.

The computational experiments were carried out on an Intel i7-10870H CPU\@ 2.20 GHz with 32 GB of memory laptop, using Windows 11 as operating system.
The programming platform was Java 11. 
Out of 32 GB, we allotted 24 GB to the Java Virtual Machine (JVM), in order to prevent, as much as possible, the interference of the garbage collector.
Details about JVM specifications, memory allocation and deallocation on the Java platform and the garbage collection process can be found in \cite{lindholm:2014}, \cite{grgic:2018}.

Prior to the metered test execution, we "warmed up" the Java Virtual Machine by performing the same test, on a smaller scale, a repeated number of times. 
This ensures that the required classes are all loaded into memory and triggers the Java-In-Time (JIT) compiler which optimizes the performance of the application by translating the bytecode to native code. 
More about Java performance techniques can be found in \cite{hunt:2016}, \cite{oaks:2014}.

We measured the running time by making the difference between the system time after and before the execution of a specific test. The space requirement
was determined by computing the application used memory before and after the test. We "suggested" the garbage collector to run before the test, to minimize its interference. 
We did not use any microbenchmark tools since the used memory increase was consistent on repeated runs and a clear indication of the memory requirement for each test. 
The source code is presented below:

{\scriptsize
\begin{verbatim}
    System.gc();
    Runtime runtime = Runtime.getRuntime();
    long usedMemoryBefore = runtime.totalMemory() - runtime.freeMemory();
    long initialTime = System.currentTimeMillis();
    test.run();
    long runningTime = System.currentTimeMillis() - initialTime;
    long usedMemoryAfter = runtime.totalMemory() - runtime.freeMemory();
    long memoryIncrease = usedMemoryAfter - usedMemoryBefore;
\end{verbatim}
}

\subsection{Creation}
The first test proves the overhead of using objects in order to represent graph vertices. Network analysis often deals with sparse graphs having tens or hundreds of millions of vertices. 
We have created empty graphs with the number of vertices ranging from $5$ to $50$ millions. 
The results are presented in Figure \ref{fig:EmptyGraph}.

\begin{figure}[htb!]
\begin{center}
\includegraphics[scale=0.5]{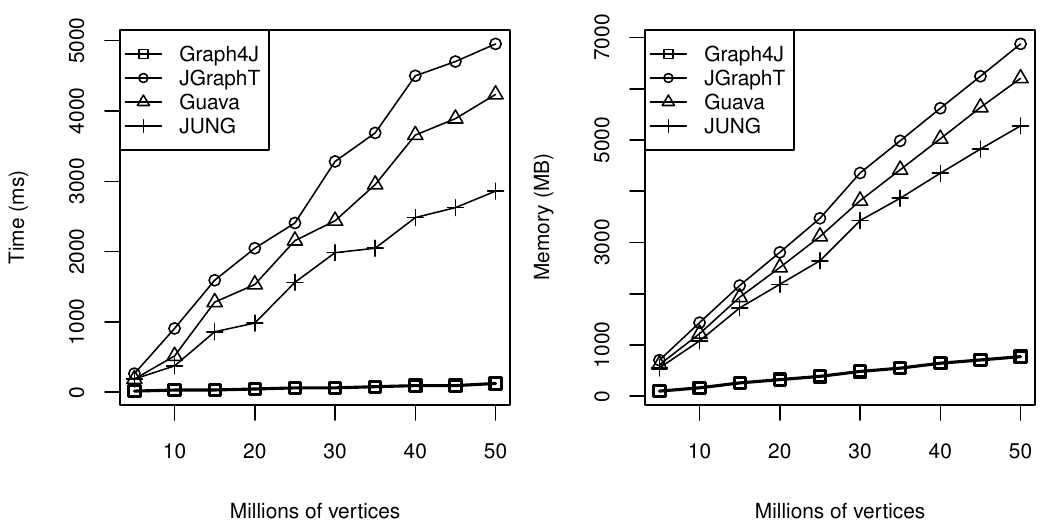}
\caption{Empty graph creation. Both the execution time and the memory requirement reflect the overhead of using objects in order to represent the vertices of a graph.}
\label{fig:EmptyGraph}
\end{center}
\end{figure}

For $n=50$ million vertices, Graph4J requires $772$ MB which is consistent with our $16n$ estimation resulting from Table \ref{table:graphMembers}.
JGraphT takes almost $9$ times more space, using $6876$ MB. As expected, Guava and JUNG require almost the same amount of memory as JGraphT, since the vertices are also {\tt Integer} objects. 
Creating collections that store objects, instead of primitive arrays, is also more expensive with respect to the running time. 
Graph4J needs only $125$  milliseconds to create the empty graph, while the other libraries need $3$ to $5$ seconds.

The second test  analyzes the case of creating complete graphs, as an indication of the general behavior regarding dense graphs. 
The difficulties here come from the large number of edges and the constraints that must be checked when adding them to the graph. 
All libraries make the difference between simple graphs, multigraphs and pseudographs, Graph4J, JGraphT, JUNG through dedicated types, 
while Guava uses builder parameters that prevent adding multiple edges between two vertices or creating self-loops. 
This type of validation is important when reading graphs from external sources, since they can expose invalid data.
Therefore, when dealing with simple graphs, in the general case, each time an edge is added we have to verify that it is valid, and this may prove costly.

The results, for a number of vertices ranging from $500$ to $5000$, are presented in Figure \ref{fig:CompleteGraph1}.

\begin{figure}[htb!]
\begin{center}
\includegraphics[scale=0.5]{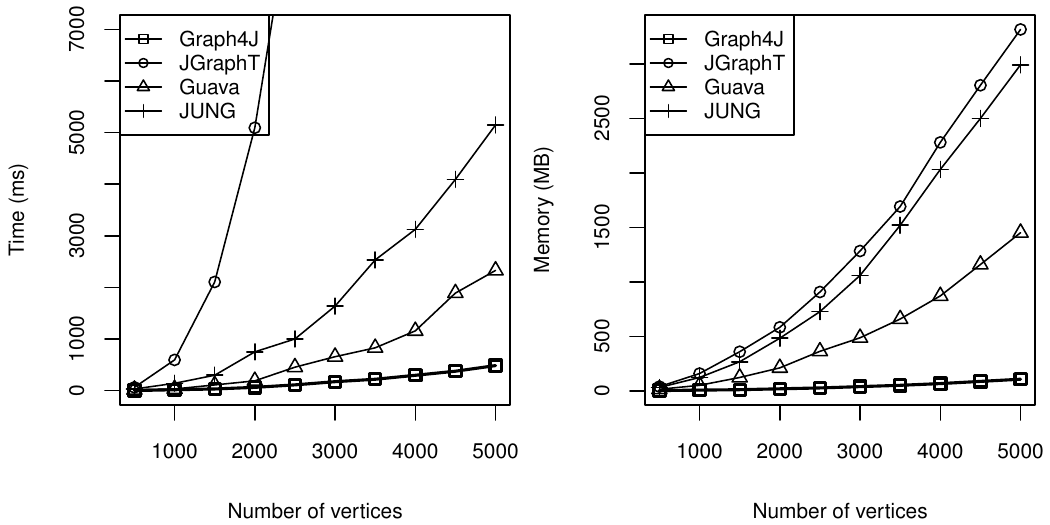}
\caption{Complete graph creation. The time required to create a simple graph depends on the efficiency of testing vertex adjacency. The used memory is highly increased if the edges are stored as objects.}
\label{fig:CompleteGraph1}
\end{center}
\end{figure}

For $n=5000$, JGraphT needs $83$ seconds and $3316$ MB of memory, compared to $484$ milliseconds and $108$ MB in our case. 
Actually, in the same amount of time we were able to create a complete graph with $40,000$ vertices and $799,980,000$ edges.
Guava was more efficient, however it still uses ten times more memory and is five times slower.
As mentioned in section \ref{dataStructures}, Graph4J has an adaptive structure based on a {\tt Bitset} that checks if two vertices are adjacent.  
It is designed to activate automatically for each vertex separately, whenever its degree exceeds an established threshold, such as in this scenario.
The other libraries do not specify explicitly the default behavior.
Although not very well documented, JGraphT seems to allow specific strategies that can be used in such cases 
({\tt FastLookupGraphSpecificsStrategy}). 
In order to obtain better results, we have tried the implementation offered by the alternative class (found in an optional package) {\tt FastutilMapIntVertexGraph}, which has a parameter that enables such a fast lookup.  
However, the running times did not improve compared to the default implementation {\tt SimpleGraph}.
The memory used by Graph4J is consistent with the $16n+17m$ (including the adjacency bitset) estimation resulting from Table \ref{table:graphMembers}.

To complete the tests regarding graph creation, we have generated large sparse $2k$-regular graphs with a fixed number of vertices, varying the degrees.
For each vertex $v$, we added edges connecting $v$ to  $u=(v + j + 1) \mod n$, where $n$ is the graph order and $j$ ranges from $0$ to $k$.
In this case, the adjacency bitsets will not be activated, due to the low number of neighbors.
The results are presented in Figure \ref{fig:SparseGraph} and
they reflect the progressive performance penalty induced by the increase in the number of edges.
Guava performs better than in the previous test, but it is still $3$ times slower and it uses $6$ times more memory.

\begin{figure}[htb!]
\begin{center}
\includegraphics[scale=0.5]{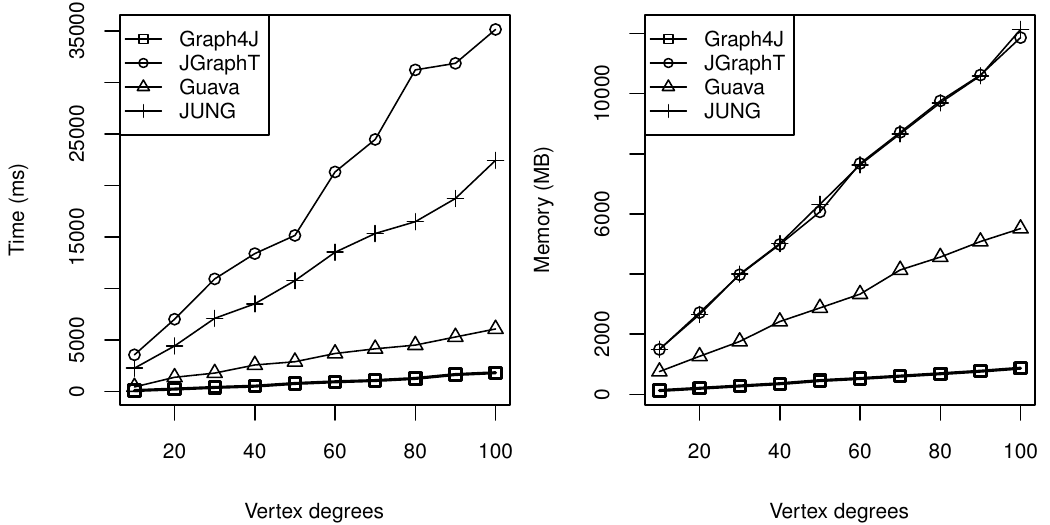}
\caption{Sparse regular graph creation. For a fixed number of $1,000,000$ vertices we varied the vertex degree of a sparse regular graph. The results reflect the penalty induced by the increase in the number of edges.}
\label{fig:SparseGraph}
\end{center}
\end{figure}

\subsection{Labels and Weights}

As we stated in section \ref{architecture}, when describing our system architecture, the way in which a graph is created starting from user objects is different in Graph4J, compared to the others. 
We consider vertices as primitive numbers and associate the user objects to them as {\it labels}, while for the other libraries the user objects {\it are} the vertices.
Regarding the edges, Guava has two options: {\tt ValueGraph} in which edges have values (objects) associated to them, just like in our case and {\tt Network} in which graph edges are user objects. 
JGraphT and JUNG follow the second approach, edges being domain specific objects. JGraphT also offers a {\tt DefaultEdge} class for the case when edges are simply connections between vertices.

The recommended usage of our library is the one described by the flow in Figure \ref{fig:Flow}. The user objects should be indexed and their indices will become the vertices of the graph. 
If the objects are globally available, it is not necessary to attach them as labels, otherwise it is possible, with a moderate memory overhead. 

Graph4J and JGraphT have both support for weighted edges, declared using the primitive {\tt double} type. In Guava and JUNG, the weight must be embedded in the user object associated with an edge. 
%Graph4J is the only one supporting weights for the vertices.

In the following test (Figure \ref{fig:LabeledGraph}) we create, on one hand (left), complete graphs starting from objects of type {\tt City} and {\tt Road} (vertices are cities and edges are roads). 
On the right, we associate weights to all edges of a complete graph.
In both cases, we measure the used memory in order to determine the impact of the structures responsible with storing the additional data.

\begin{figure}[htb!]
\begin{center}
\includegraphics[scale=0.5]{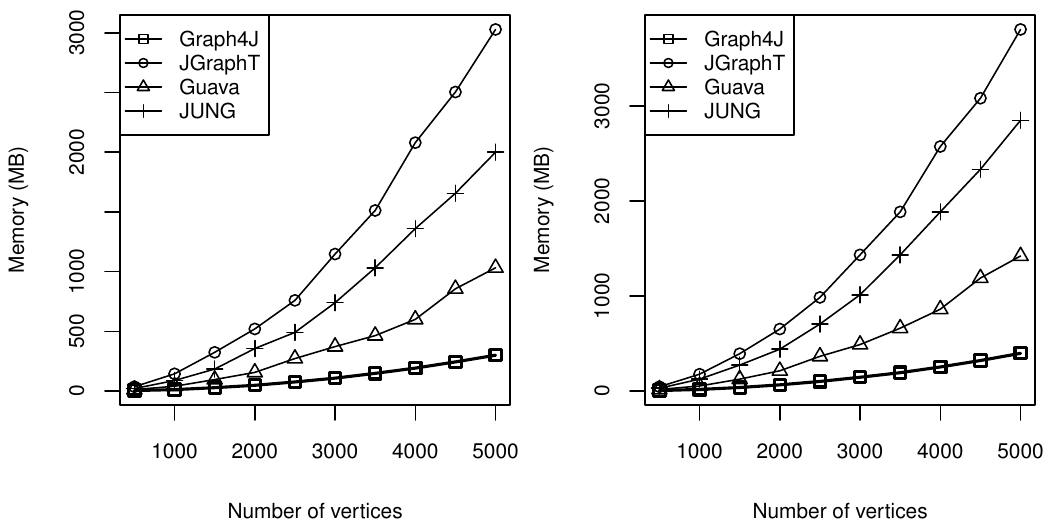}
\caption{Memory requirement of labeled complete graphs having user objects associated with both vertices and edges (left) 
and complete weighted graphs, having weights associated with their edges (right).}
\label{fig:LabeledGraph}
\end{center}
\end{figure}

For the labeled graphs, Graph4J needs $4n + 8m$ bytes for storing the references to the objects (see Table \ref{table:graphMembers}), so the total requirement is $20n + 25m$, including the adjacency bitset. 
For $n=5000$, the reported value is $300$ MB, close to our $298$ MB estimation. Note that the memory occupied by the {\tt City} and {\tt Road} objects was not taken into consideration.
In the edge weighted case, we need $4n + 16m$ for storing weights ({\tt double} is represented on $8$ bytes) so the total is $20n + 33m$, including the adjacency bitset. 
For $n=5000$, the reported value is $396$ MB, and our estimation $393$ MB.
We also performed this test using Guava's {\tt Network} type, instead of {\tt ValueGraph} and the memory requirement increased above the one of JUNG.

\subsection{Alteration}
The next tests were performed on graphs already created, in order to highlight the performance of a specific operation, without the overhead of the creation process. 
Graph alteration may seem a less important operation, but it is present in the implementation of various algorithm. 
We used it when determining the eulerian circuit of a graph, where we created a copy of the original graph and kept removing edges out of it as we added them to the eulerian circuit being constructed.

Except Graph4J, all the other libraries do not seem to allow the alteration of the graph while iterating over the vertex or edge sets returned by their specific API, so we made copy of them before the removal process. 
The overhead of creating these copies was also not measured.
The edge removal test (Figure \ref{fig:RemoveEdgesNodes}) was performed on complete graphs with the number of vertices ranging from $200$ to $2000$, 
while the vertex removal was carried out on random sparse graphs with $10,000$ to $100,000$ vertices having an average degree of $100$. We generated the sparse graphs using the {\it Gnm} Erd\"{o}s-R\'enyi model, 
which ensures a random uniform selection from the set of all possible graphs with $n$ nodes and $m$ edges.

\begin{figure}[htb!]
\begin{center}
\includegraphics[scale=0.5]{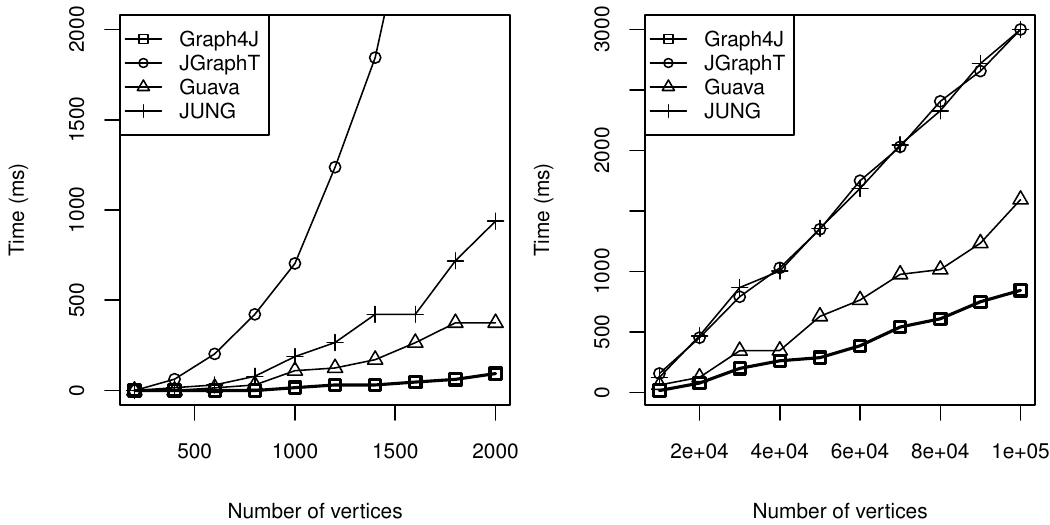}
\caption{Time required by operations that alter a graph: removing all edges from a complete graph, one by one (left) and removing all nodes from a random sparse graph, one by one (right).}
\label{fig:RemoveEdgesNodes}
\end{center}
\end{figure}

For Graph4J, removing edges while iterating over them offers the best performance since the operation is performed in constant time.
If we had removed all these edges in a random fashion, the time complexity  for removing an edge incident with $v$ would have been $O(d_G(v))$.
In this case, the response times of Graph4J and Guava are almost the same.
%Graph4J, both having better response times than JGraphT and JUNG. 

\subsection{Inspection}
In this test we create random tournaments, which are directed graphs obtained by orienting all the edges of a complete graph, iterate over all their vertices and, 
for each vertex, iterate over all its predecessors and successors.
Since most algorithms rely on inspecting the adjacency lists, the ability to perform this iterations efficiently is important for their overall performance.
As in the previous case, the graphs were created before the metered tests.
To better capture the behavior of the graph implementations, we repeated the iteration process $10$ times and Figure \ref{fig:Iterate} contains the total cumulative time.

\begin{figure}[htb!]
\begin{center}
\includegraphics[scale=0.5]{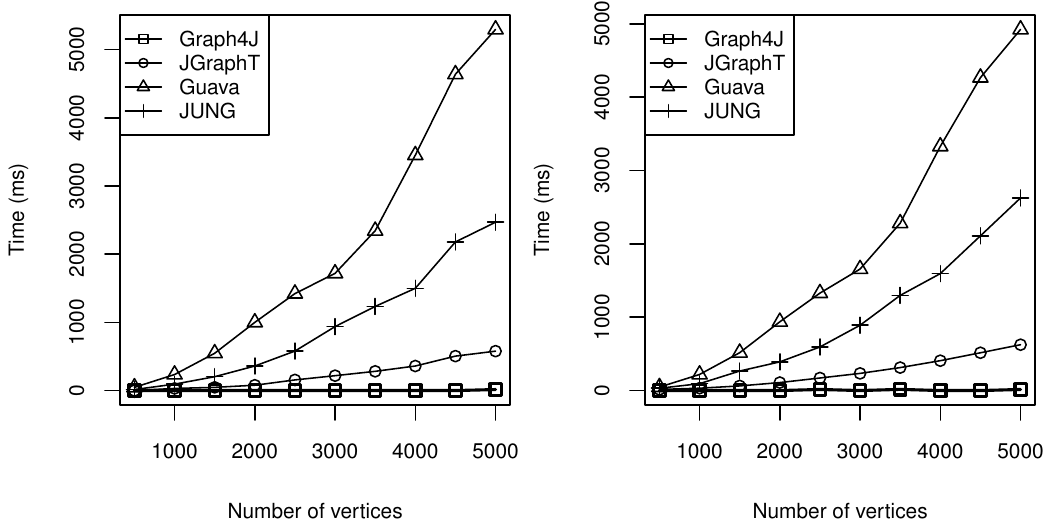}
\caption{Time required by operations that inspect a graph: iterating over the successors of each vertex (left) and iterating over the predecessors of each vertex (right).}
\label{fig:Iterate}
\end{center}
\end{figure}

Due to the nature of our graph representation, the fact that Graph4J performs the iterations very quickly (less than $15$ ms) is somehow expected, while JGraphT is close, finishing in less than $1$ second. 
Guava needs around $5$ seconds for the largest instance, which may be the penalty for an infrastructure supporting more elaborate ways of iterating through adjacency lists.

\subsection{Traversal}
Graph traversal is the basis of many algorithms, the most common methods being depth first search (DFS) and breadth first search (BFS).
Except for JUNG, all other libraries offer support for performing these types of traversals in a standard manner.
JGraphT uses the {\it iterator} pattern, providing the classes {\tt DepthFirstIterator} and {\tt BreadthFirstIterator}, which return, one by one, the vertices of the graph according to the traversal type.
Guava offers the {\tt Traverser} class, containing methods such as {\tt depthFirstPostOrder}, {\tt depthFirstPreOrder} and {\tt breadthFirst}.
They use the {\it visitor} pattern which aims at separating the traversal algorithm from the action that must be executed when a vertex of a graph is visited.
Graph4J implements both patterns. It contains the classes {\tt DFSIterator} and {\tt BFSIterator} which returns the graph vertices as objects of type {\tt SearchNode}, 
including additional information related to the corresponding DFS or BFS tree: the level of the node, its parent, visiting time and the connected component it belongs to.
The visitor pattern is implemented by the classes {\tt DFSTraverser} and {\tt BFSTraverser} which receive as argument either a {\tt DFSVisitor} or {\tt BFSVisitor}, 
defining the methods that will be invoked during the search, such as {\tt startVertex}, {\tt finishVertex}, {\tt treeEdge}, {\tt backEdge}, etc.
\begin{comment}
The {\tt BFSVisitor} contains the following methods that will be invoked by the search algorithm: 
{\tt startVertex}, {\tt finishVertex}, {\tt treeEdge}, {\tt backEdge}, {\tt forwardEdge}, {\tt upward}, depending on the graph type being explored,
while the {\tt BFSVisitor} contains the methods: {\tt startVertex}, {\tt finishVertex}, {\tt treeEdge}, {\tt backEdge}, {\tt crossEdge}.
The visiting algorithm can be stopped from any of these methods, using the {\tt interrupt} method.
\end{comment}

In order to test the performance of the traversers, 
we generated random graphs using the {\it Gnp} Erd\"{o}s-R\'enyi model
in which, for a graph with $n$ vertices, edges are chosen with a probability of $p$. We used $p=0.2$ and $n$ ranging from $100$ to $1000$.
As in the previous cases, we created the graphs before the actual tests and we measured only the running times and the additional memory required by the search algorithms. 
Within a test, for a given graph, we performed $n$ traversals, starting from each of its vertices.
For Graph4J, we have used iterators.
The results obtained for the depth first search are presented in Figure \ref{fig:DFSIterator}.

\begin{figure}[htb!]
\begin{center}
\includegraphics[scale=0.5]{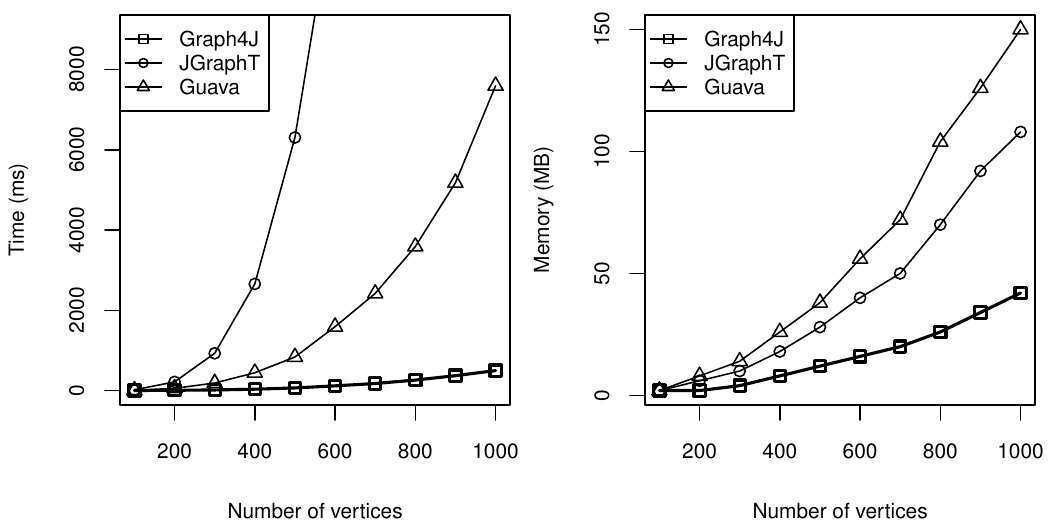}
\caption{The running time and the additional memory required by DFS traversals of random {\it Gnp} graphs with $p=0.2$}
\label{fig:DFSIterator}
\end{center}
\end{figure}

Despite the fact that during the search algorithm Graph4J creates objects of type {\tt SearchNode}, containing additional information about each visited vertex, there is a major gap between our running times and the others.
For $1000$ vertices, Graph4J finishes in about $500$ milliseconds, while Guava needs $7.5$ seconds and JGraphT $95$ seconds.
Regarding the occupied memory, Graph4J uses an additional $42$ MB in order to create the search nodes, as it iterates over the graph. 
The other libraries do not provide additional information and still use more memory, up to $150$ MB.

The BFS traversal (Figure \ref{fig:BFSIterator}) follows a similar pattern.
The running times of Graph4J are an order of magnitude better, while the memory consumption is slightly increased due to the nature of breadth first search.

\begin{figure}[htb!]
\begin{center}
\includegraphics[scale=0.5]{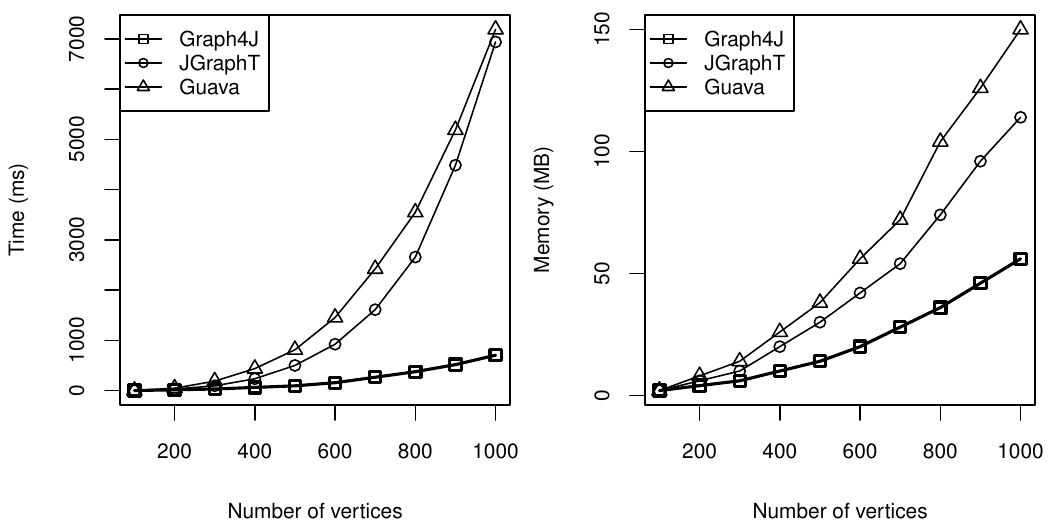}
\caption{The running time and the additional memory required by BFS traversals of random {\it Gnp} graphs with $p=0.2$}
\label{fig:BFSIterator}
\end{center}
\end{figure}

\subsection{Algorithms}
In this subsection, we compared algorithms from Graph4J against JGraphT. Guava does not provide any classical algorithm implementation, 
while JUNG is mainly focused on data mining and social network analysis.

The comparison uses the following algorithms: Dijkstra's
shortest path, Prim's and Kruskal's minimum spanning tree, Edmnonds-Karp maximum flow and Hopcroft-Karp maximum cardinality matching.
A detailed description of the algorithms we have tested can be found in \cite{cormen:2009}, \cite{algs4:2011}. 
Unless otherwise specified, in the following paragraphs $n$ represent the number of vertices and $m$ the number of edges of the tested graphs.

Dijkstra's algorithm aims at finding the shortest paths in a directed weighted graph, from a single source vertex to all other vertices.
It is applicable only if all the edge weights are non-negative.
We tested the performance of the implementations both for sparse and dense graphs, for each one of them starting the algorithm repeatedly, with the source in every vertex, as if we were determining all pairs shortest paths. 
The sparse graphs were generated using the {\it Gnm} model, with $n$ ranging from $50,000$ to $500,000$ and an average vertex degree of $50$.
The dense graphs were created using the {\it Gnp} model, the edge probability being $0.5$. 
JGraphT implementation uses a Fibonacci heap having a time complexity of $O(m + n \log n)$, while Graph4J a binary heap, the complexity  being $O(m \log n)$. 
%The  test has the time complexity $O(n(m + n \log n))$.
As in the previous cases, the time required to create the graphs was not measured. The results are presented in Figure \ref{fig:Dijkstra}.

\begin{figure}[htb!]
\begin{center}
\includegraphics[scale=0.5]{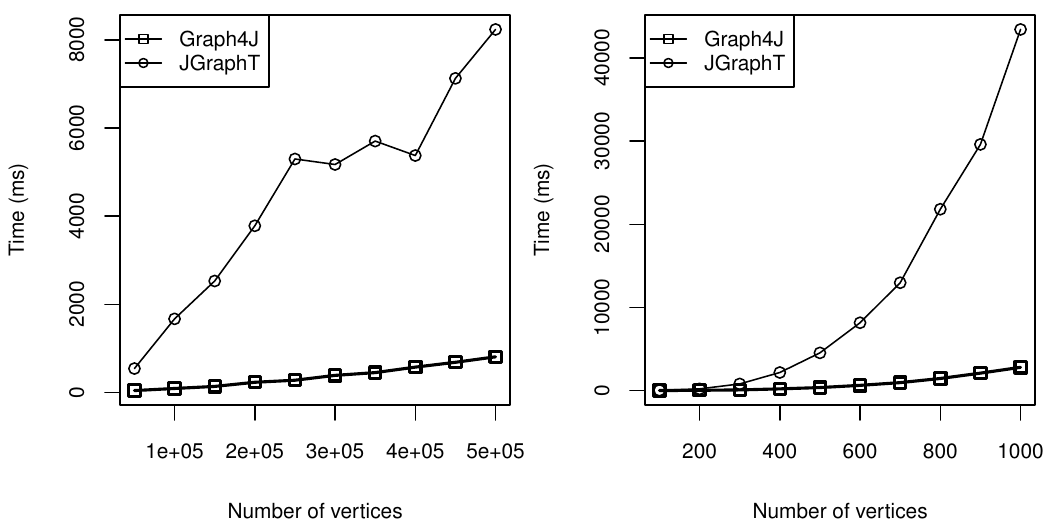}
\caption{The running times required to determine all pairs shortest paths using Dijkstra's algorithm, on random sparse graphs with an average degree of $50$ (left) and on random dense graphs with an edge probability of $0.5$ (right).}
\label{fig:Dijkstra}
\end{center}
\end{figure}

In both cases, Graph4J runs an order of magnitude faster than JGraphT,
despite the fact that, theoretically, the Fibonacci heap is more efficient than a standard binary heap. 
Being a state-of-the-art library, JGraphtT's algorithm implementations are very well crafted and tested. We conclude that the
difference in performance is due to the graph infrastructure, our lower-level approach being substantially more efficient than the object-oriented one.
JUNG also contains an implementation of this algorithm, but its runtime performance was not so good, so we didn't include it in our test.
A comparison between JGraphT and JUNG can be found in \cite{jgrapht}.

We also tested Prim's and Kruskal's minimum spanning tree algorithms.
The Prim's algorithm implementation is similar to the Dijkstra's shortest path and the end results followed the same pattern as in Figure \ref{fig:Dijkstra}.
In the case of Kruskal's, it is exposed another aspect of our graph structure. Since the algorithm starts by sorting the edges by their weight, we need to explicitly create all those edges and store them in a collection. 
By default, the {\tt Edge} objects do not exist in our graph structure, as they are created dynamically whenever they are needed.
This will drastically increase the required memory, especially for graphs with many edges. 
For this type of algorithms, the space efficiency of our library decreases, but it still remains superior if we were to add the total amount of memory required also by the graph itself.
The time complexity of the test is $O(m \log m)$, or equivalently, $O(m \log n)$.
The results in Figure \ref{fig:Kruskal} show that the running time is up to $5$ times better, but the memory needed explicitly for running the algorithm is much higher.

\begin{figure}[htb!]
\begin{center}
\includegraphics[scale=0.5]{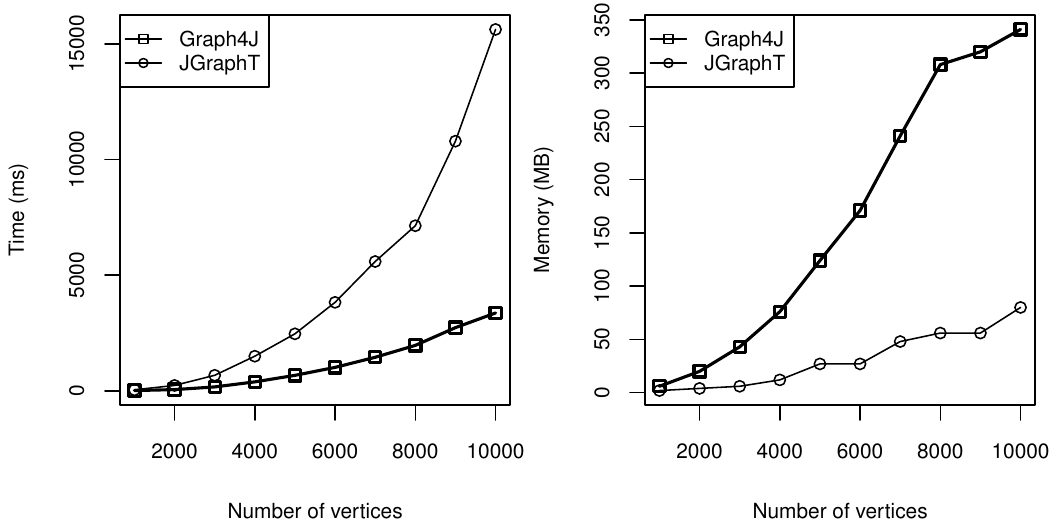}
\caption{The running times and the additional memory required by Kruskal's minimum spanning tree algorithm, for random {\it Gnp} graphs with $p=0.1$.}
\label{fig:Kruskal}
\end{center}
\end{figure}

The next algorithm we tested was Edmonds-Karp, for computing the maximum flow. The particularity in this case is that we have to store for each edge its flow. 
In JGraphT, the flow is represented as a {\tt Map<E, Double>}, where $E$ is the edge type. 
We used a structure that does not require creating the edge objects, similar to those used for edge weights and labels (see Figure \ref{fig:GraphStructure}).
Since our successor and predecessor iterators return not only the vertex but also its position in the adjacency list, we can access the flow data in $O(1)$ time while iterating, with much less memory than a hashtable. 
The complexity of the algorithm is $O(n m^2)$.
We performed the tests on randomly generated networks, using the {\it Gnp} model, with $n$ ranging from $500$ to $5000$ and the edge probability $0.2$.

\begin{figure}[htb!]
\begin{center}
\includegraphics[scale=0.5]{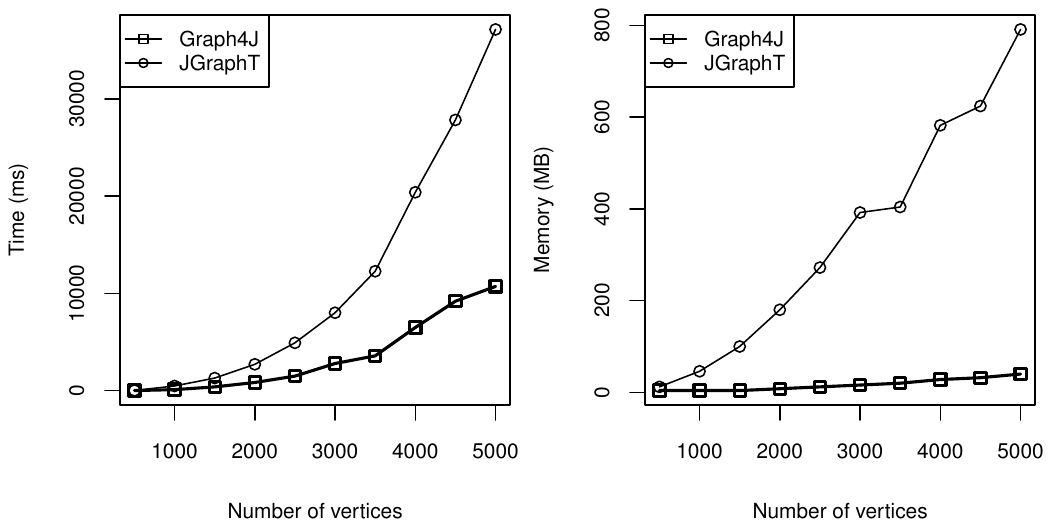}
\caption{The running times and the additional memory required by Edmonds-Karp maximum flow algorithm, for random {\it Gnp} networks with $p=0.1$.}
\label{fig:EdmondsKarp}
\end{center}
\end{figure}

This low-level approach pays off, the results in Figure \ref{fig:EdmondsKarp} show running times up to $3$ times smaller, with very little additional space required.

Finally, we present the test for the Hopcroft-Karp algorithm, for computing the maximum cardinality matching in a bipartite graph.
We did not follow the method presented in \cite{jgrapht} (where JGraphT implementation is compared against other libraries) that creates random bipartite graphs which are guaranteed to contain a perfect matching. 
Instead, we generate completely random bipartite graphs using the {\it Gnp} model, with $n$ ranging from $1,000$ to $10,000$ and $p=0.1$ (each partition set contains $n/2$ vertices). 
Both libraries were required to determine first the two classes of the bipartite graph and then to find the maximum cardinality matching.
The time complexity of the test is $O(n+m)$ for determining the bipartition plus $O(m \sqrt n)$ for the Hopcroft-Karp algorithm.
The results are presented in Figure \ref{fig:HopcroftKarp}.

\begin{figure}[htb!]
\begin{center}
\includegraphics[scale=0.5]{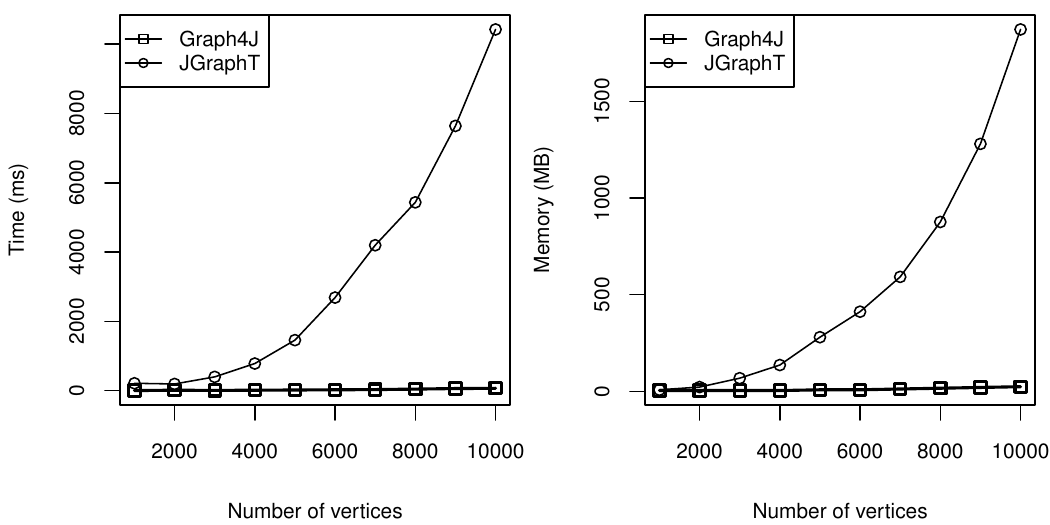}
\caption{The running times and the additional memory required by Hopcroft-Karp maximum cardinality matching algorithm, for random {\it Gnp} bipartite graphs with $p=0.1$.}
\label{fig:HopcroftKarp}
\end{center}
\end{figure}

The gap between the two implementations is large this time, for $n=10,000$ vertices Graph4J finds the optimum matching in $65$ milliseconds while JGraphT needs more than $10$ seconds. 
We also tested denser graphs, using $p=0.5$, as the authors of JGraphT mentioned in \cite{jgrapht} that on sparse graphs the performance deteriorates, but the results followed the same pattern.
A possible explanation could be the fact that we use our specialized collections, {\tt VertexQueue} in the BFS phases of the algorithm and {\tt VertexStack} in the DFS phases, instead of standard Java collections. 
We assume that the usage of the object-oriented collections takes a heavy toll not only on the running time, but also on the used memory: more than $1.5$ GB compared to $42$ MB.

\section{Conclusions}
\label{conclusions}
In this paper we have presented the design and the performance of the Graph4J library, 
aimed at offering support for solving problems that can be modeled using graphs and require graph related algorithms, on Java programming platform.
Taking into consideration the fact that, when we started working at this project, there were three well established graph libraries JGraphT, JUNG and Guava, 
we did not particularly want to compete with them but to offer an alternative approach to the object-oriented model used by them,
especially suitable for problems that use a simple mathematical formulation.
By representing the graph internals as array-based primitive data structures, we obtained significant improvements regarding both the required memory and the running times of the algorithm implementations.
Future work will follow in several directions. An immediate goal is to add more fundamental algorithm implementations, in order to cover the basic necessities of a project requiring graph related support. 
To increase interoperability with other software solutions, we want to cover most of the popular graph specification formats, not only the basic ones  found in the initial release.
We also plan to introduce packages related to network analysis, data mining and layout algorithms for graph visualization.
To become a viable alternative to existing libraries, we intend to attract a larger community of users and developers 
by providing an extensive documentation of the programming interface and by using it in the process of teaching graph algorithms to students.

\bibliographystyle{apalike}
\bibliography{graph4j}
\end{document}